\RequirePackage{lineno}  
\documentclass[11pt,a4paper]{article}
\usepackage{jheppub}

\usepackage{graphicx}
\usepackage{amssymb,amssymb}
\usepackage{lineno}
\usepackage{atlasphysics}
\usepackage{subfigure}
\usepackage{mathrsfs}
\usepackage{slashed}
\usepackage{setspace}
\usepackage{amsmath}
\usepackage{multirow}
\usepackage{chngpage}
\usepackage{url}
\usepackage{mciteplus}

\newlength{\figwidth}
\newcommand{\pmasym}[2]{^{+#1}_{-#2}}
\newcommand{\sigmattbar}{\ensuremath{\sigma_{\ttbar}}}

\newcommand{\lumitot}{\mbox{0.70~fb$^{-1}$}}
\newcommand{\lumitotpm}{\mbox{0.70$\pm$0.03~fb$^{-1}$}}

\newcommand{\xsecctot}{176}
\newcommand{\xseccstat}{\pm 5}
\newcommand{\xseccsyst}{{^{+14}_{-11}}} 
\newcommand{\xsecclumi}{\pm 8}

\newcommand{\AKT}{\rm{anti}-k$_{t}$}

\def\MCatNLO{{\sc MC@NLO}}

\def\Zg{\Zboson/\gamma^{*}}
\def\GEANT{{\sc GEANT4}}

\begin{document}

\title{\vspace{-0.5cm} \flushright{\normalsize{CERN-PH-EP-2011-223}}
  \\ \center{
Measurement of the cross section for top-quark
pair production in $pp$\  collisions at $\sqrt{s}=7$~TeV with the
ATLAS detector using final states with two high-$\pt$ leptons
}
}

\author{The ATLAS Collaboration}

\abstract{
 A measurement is reported of the production cross section of
top-quark pairs ($\ttbar$) in proton-proton collisions at a
center-of-mass energy of 7~\TeV\ recorded with the ATLAS detector at
the LHC.
Candidate events have a signature
consistent with containing two isolated leptons, large missing
transverse momentum, and at least two jets. Using a data sample
corresponding to an integrated luminosity of \lumitot, a \ttbar\
production cross section $\sigmattbar=\xsecctot \xseccstat ({\rm
stat.})\xseccsyst ({\rm syst.}) \xsecclumi ({\rm lum.})~{\rm pb}$ is
measured for an assumed top-quark mass of $m_t=172.5~\GeV$.
This measurement is in good agreement with
Standard Model predictions.
}

\date{today}


\maketitle

\section{Introduction}\label{s:intro}

As the heaviest known elementary particle, the top quark is a
particularly interesting probe of the Standard Model (SM). The
measurement of the $\ttbar$ production cross section in different
decay modes is a sensitive test of perturbative QCD and the SM
description of top-quark decay. The production cross section in
proton-proton ($pp$) collisions at a center-of-mass energy
$\sqrt{s}=7~\TeV$\ is calculated to be $165\pmasym{11}{16}$~pb at
approximate next-to-next-to-leading-order (NNLO)~\cite{topxs,Langenfeld:2009tc}, and the top quark
is predicted to decay nearly 100\% of the time to a $W$ boson and a
$b$ quark.  A measured cross section that differs from
the SM prediction can be a sign of new physics.  Furthermore,
$\ttbar$ production is an important background in many searches for
physics beyond the SM, and in searches for the SM Higgs boson.

The $\ttbar$ event topologies are determined by the
decays of the two $W$ bosons.  In increasing order of $\ttbar$
branching fraction: dilepton final states occur when both $W$ bosons
decay to a charged lepton  and a neutrino,
`lepton plus jets' final states when only one $W$ boson decays
leptonically while the other decays to a pair of quarks, and
all-hadronic final states when both $W$ bosons decay to pairs of
quarks.

Top-quark production in dilepton final states has been studied using
proton-antiproton collisions at
$\sqrt{s}=1.96~\TeV$~\cite{Aaltonen:2010bs,Abazov:2009ae} and
Large Hadron Collider (LHC)
measurements of the production cross section in proton-proton
collisions at $\sqrt{s}=7~\TeV$ in the same final state have recently been
reported~\cite{Chatrchyan:2011nb,ATL-CONF-2011-034}.
A measurement is presented of the \ttbar\ production cross section
using the dilepton channel, characterized by two opposite-sign
leptons, unbalanced transverse momentum indicating the presence of
neutrinos from the $W$-boson decays, and two $b$-quark jets. This
result uses twenty times more data than the previous ATLAS measurement in the same final state,
reported in Ref.~\cite{ATL-CONF-2011-034}.

The \ttbar~dilepton final states can be selected with a good
signal-to-background ratio using simple kinematic requirements.  With
the additional requirement of the
presence of a jet consistent with a $b$~quark
(`$b$-tag'), the signal-to-background ratio can be further improved.
Cross-section measurements with and without the $b$-tag requirement
are reported here.
 Leptons are either well-identified electron or
muon candidates that are selected using the full detector or, to
reduce losses from lepton identification inefficiencies, isolated
tracks.  The well-identified electrons or muons are called
`identified leptons', and the isolated tracks are referred to as
`track leptons'. The term `lepton' is used to refer to identified
leptons and track leptons collectively. Events with one identified
lepton and one track lepton are called `lepton+track' events.
  Each dilepton channel is
exclusive, i.e. has no overlap with the other channels. Channels
with tau leptons are not explicitly reconstructed, but reconstructed
leptons can arise from leptonic tau decays and a track lepton can
arise from hadronic tau decay modes as well.  The analysis with the
$b$-tag requirement uses only identified leptons.

The measured cross section takes into account the $\ttbar$\ signal
acceptance and the expected background contributions from
$\Zg$+jets, single top quarks, $WW$, $WZ$, and $ZZ$ events, and
events with misidentified leptons (primarily $W+$jets events).
Background contributions from
$\Zg\rightarrow ee$+jets, $\Zg\rightarrow\mu\mu$+jets and events with
misidentified leptons are evaluated
directly from the data. All other background contributions are
evaluated using Monte Carlo (MC) simulation samples.

\section{Detector and data sample}
\label{s:detector}

The ATLAS detector~\cite{atlasdet} at the LHC covers nearly the
entire solid angle around the collision point. It consists of an
inner tracking detector (ID) comprising a silicon pixel detector, a
silicon microstrip detector (SCT), and a transition radiation
tracker. The ID is surrounded by a thin superconducting
solenoid providing a 2~T magnetic field, and by liquid-argon
electromagnetic sampling calorimeters (LAr) with high granularity.
An iron-scintillator tile calorimeter provides hadronic energy
measurements in the central pseudorapidity\footnote{In the
right-handed ATLAS coordinate system, the
  pseudorapidity $\eta$ is defined as $\eta=-\ln[\tan(\theta/2)]$,
  where the polar angle $\theta$ is measured with respect to the LHC
  beamline. The azimuthal angle $\phi$ is measured with respect to the
  $x$-axis, which points towards the centre of the LHC ring.  The
  $y$-axis points up.  Transverse momentum and energy are defined as
  $\pT = p\sin\theta$ and $\ET = E\sin\theta$, respectively.}\ range ($|\eta|< 1.7$).
The end-cap and forward regions are instrumented with LAr
calorimetry for both electromagnetic (EM) and hadronic energy
measurements up to $|\eta|< 4.9$. The calorimeter system is
surrounded by a muon spectrometer incorporating three
superconducting toroid magnet assemblies, with
bending power between 2.0 and 7.5 Tm.

A three-level trigger system is used to collect data. The
first-level trigger is implemented in hardware and uses a subset of
the detector information to reduce the rate to at most 75\,kHz. This
is followed by two software-based trigger levels that together
reduce the event rate to $\sim 300$ Hz.

The analysis uses collision data with a center-of-mass energy of
$\sqrt{s}$ = 7~\TeV\ recorded in 2011, with an integrated luminosity
of \lumitotpm~\cite{lumiPub, lumi}.

\section{Simulated samples}
\label{s:mc}

Monte Carlo simulation samples are used to calculate the $\ttbar$\
acceptance and to evaluate the background contributions from single
top quarks, $WW$, $WZ$, and $ZZ$ events, and $\Zg
\rightarrow\tau\tau$+jets.
All MC samples are processed with the
\GEANT~\cite{geant4}\ simulation of the ATLAS detector~\cite{atlsim}\
and events are passed through the same analysis chain as the data.

The generation of \ttbar{} and single top-quark events uses the
\MCatNLO\ generator~\cite{mcatnlo1,Frixione:2003ei,Frixione:2005vw}\
with the CTEQ6.6~\cite{cteq66}\ parton distribution function (PDF)
set and a top-quark mass of 172.5~\GeV. Expected $\ttbar$\ yields
are calculated with a cross section normalized to the prediction of
{\sc Hathor}\cite{Aliev:2010}, which employs an NNLO perturbative QCD calculation.
Single top-quark production with \MCatNLO\ includes the $s$, $t$ and
$Wt$ channels and the diagram-removal scheme~\cite{diagrem}\  is
used to reduce overlap with the $\ttbar$\ final state.

Drell-Yan events ($\Zg$+jets) are modeled with the {\sc Alpgen}
generator, using the MLM matching scheme~\cite{alpgen}\ and the
CTEQ6L1 \cite{cteq6l}\ PDF set. The $\Zg$+jets samples, including
both light and heavy flavor jets, are normalized to NNLO with a
$K$-factor of 1.25. 
In the $\Zg\rightarrow ee$ and $\mu\mu$ decay channels, the
background from $\Zg$+jets is evaluated using a data-driven
technique that normalizes the MC expectation to the data observation
near the $Z$ pole. Background contributions from the $W$+jets final
states come primarily from events where the $W$\ boson decays
leptonically and the second lepton candidate is a misidentified jet
or a heavy-flavor decay. Backgrounds from $W$+jets 
events are evaluated from the data.

All MC simulated events are hadronized using the {\sc Herwig}\ shower
model~\cite{herwig1,Corcella:2002jc}\ supplemented by the
{\sc Jimmy}\ underlying event model~\cite{jimmy}.
Both hadronization programs are tuned to ATLAS data using the ATLAS
MC10 tune~\cite{ATLAS:1303025}.
Diboson events are modeled using the {\sc Alpgen}\ generator
normalized with $K$-factors of 1.26 ($WW$), 1.28 ($WZ$) and 1.30
($ZZ$) to match the total cross section from NLO QCD
predictions using calculations with the MCFM
program~\cite{Campbell:1999ah}.

All Monte Carlo samples are generated taking into account that
multiple $pp$ interactions can occur in the same LHC bunch crossing
within a given event (`pile-up').  The average number of interactions per
crossing is 5.6 in this data set. The MC events are re-weighted so
that the distribution of interactions per crossing in the MC matches
that observed in the data.

\section{Object selection}
\label{s:object}

Leptons are required to be isolated and have high transverse
momentum, $\pt$, consistent with originating from $W$-boson
decay, with $\pt$ thresholds chosen to ensure events are triggered
with high efficiency.

Electron candidates are reconstructed from energy deposits
(clusters) in the EM calorimeter, which are then associated to
reconstructed tracks of charged particles in the inner detector.
Stringent quality requirements on the conditions of the EM
calorimeter at the time of data taking are applied to ensure a well measured reconstructed
energy. A `tight' selection~\cite{ElectronPerformance} using calorimeter,
tracking and combined variables, is employed to provide good
separation between the signal electrons and background.  Electron
candidates are additionally required to have $\pt>25\GeV$ and
$|\eta_{\text cl}|<2.47$, excluding electrons from the transition
region between the barrel and endcap calorimeters defined by
$1.37<|\eta_{\text cl}|<1.52$. The variable $\eta_{\text cl}$\ is
the pseudorapidity of the energy cluster associated with
the candidate.

Muon candidate reconstruction is begun by searching for track
segments in
 layers of the muon chambers.
These segments are  combined starting from the
outermost layer, fitted to account for material effects, and matched
with tracks found in the inner detector.
The candidates are refitted using the complete track information from
both detector systems
, and required to satisfy $\pt>20\GeV$ and
$|\eta|<2.5$.

Lepton isolation requirements reduce backgrounds from misidentified
jets and suppress the selection of leptons from heavy-flavor decays.
For electron candidates, the transverse energy ($\ET$) deposited in
the calorimeter not associated to the electron is summed in a cone
 of radius $\Delta R = 0.2$, where $\Delta R\equiv\sqrt{\Delta\eta^2 + \Delta\phi^2}$, around the
electron and is required to be less than $3.5\GeV$. For muon
candidates, the isolation requirement is based on both calorimeter
and track information. The track isolation requirement is based on the
sum of the transverse momenta of tracks with $\pt>1$~GeV in a
cone $\Delta R = 0.3$\ centered on the muon candidate, while the
calorimeter isolation requirement is based on the sum of transverse
energy in the same cone.  Both the track and calorimeter sums are
required to be less than 4~GeV.
Additionally, muon candidates
must have a distance $\Delta R > 0.4$\ from any jet with
$\pt>20$~GeV, further suppressing muon candidates from heavy flavor
decays.
Muon candidates arising from cosmic rays are rejected by removing
candidate pairs that are back-to-back in the $r-\phi$ plane and
with transverse impact parameters relative to the beam axis
$|d_0|~>~0.5~$mm.

Track-lepton (TL) candidates are defined by an ID track with $\pt >
25\GeV$ and a series of quality cuts optimized for high efficiency
and a low rate of misidentification. The track must have at least
six  SCT hits  and at least one hit in the innermost pixel layer. It
also must have $|d_0| < 0.2$~mm and the uncertainty on the momentum
measurement must be less than 20\%. The track has to be isolated
from other nearby tracks, following the track isolation defined above, in this case using
 tracks with $\pt>0.5\GeV$.  The summed momentum cut is set to 2~GeV.

Jets are reconstructed with the {\it \AKT} algorithm~\cite{antikt}
with a radius parameter $R = 0.4$, starting from energy clusters
in the calorimeter reconstructed using the scale established for electromagnetic objects.
These jets are then calibrated to
the hadronic energy scale using $\pt$\ and $\eta$\ dependent
correction factors~\cite{jetcor}.
Jets are removed if they are within $\Delta R=0.2$\ of a well-identified
electron candidate or a TL.
The jets used in the analysis are required to have $\pt>25$~GeV
and $|\eta|<2.5$.

Jets are identified as $b$-quark candidates (`$b$-tagged') by an
algorithm that forms a likelihood ratio of $b$- and light-quark jet
hypotheses using the following discriminating variables: the signed
impact parameter significance of well measured tracks associated
with a given jet, the decay length significance associated with a
reconstructed secondary vertex, the invariant mass of all tracks
associated to the secondary vertex, the ratio of the sum of the
energies of the tracks associated with the secondary vertex to the
sum of the energies of all tracks in the jet assuming a pion
hypothesis, and the number of
two-track vertices that can be formed at the secondary
vertex~\cite{AdvBtag}. The cut on the combined likelihood ratio has
been chosen such that a $b$-tagging efficiency of $\approx 80\%$\
per $b$-jet in \ttbar~candidate events is achieved.

The missing transverse momentum is formed from the negative vector
sum of transverse momenta of all jets with $\pt>20$~GeV and $|\eta|
< 4.5$~\cite{METPaper}. The contribution from cells associated with
electron candidates is replaced by the candidates' calibrated
transverse energy. The contribution from all muon candidates  and
calorimeter clusters (including those not belonging to a reconstructed
object) is also included.
The symbol $\MET$ is used to denote the magnitude of the
missing transverse momentum.

\section{Event selection}
\label{s:event}

The analysis requires collision data selected by an inclusive single
electron or muon trigger with offline-reconstructed
candidates satisfying $\pt>25\GeV$ for electrons, and $\pt>20\GeV$ for muons, to ensure a constant trigger efficiency.
To ensure that the event was triggered by the lepton candidates used
in the analysis,  one of the identified leptons and the triggered
lepton are required to match within $\Delta R < 0.15$.

Events are required to have a primary interaction vertex with at
least five tracks with $\pt>400~\MeV$.  The event is discarded if
any jet with $\pt>20$~GeV fails quality cuts designed to reject jets
arising from calorimeter noise or activity inconsistent with the
bunch-crossing time~\cite{jetcor}. If an electron candidate and a
muon candidate share a track, the event is also discarded.

The selection of events in the signal region consists of a series of kinematic requirements on
the reconstructed objects.
The requirements on $\MET$, the lepton-lepton invariant
mass ($m_{\ell\ell}$), and the scalar $\pt$ sum of all selected jets and leptons ($\HT$) are optimized
to minimize the expected total uncertainty on the cross-section measurement.
The resulting event selection, referred to as the `non-$b$-tag' selection, is listed below.

\begin{itemize}

\item Events must have exactly two oppositely-charged identified-lepton candidates ($ee$, $\mu\mu$,
  $e\mu$), satisfying the selection criteria of Section~4, {\em or} if only one identified-lepton candidate is found, the event is retained if a track-lepton candidate is present, with opposite charge to the identified lepton, forming a lepton+track event ($e$TL or $\mu$TL).

\item Events must have at least two jets with $\pT>25\GeV$ and $|\eta|<2.5$.

\item Events in the $ee$, $\mu\mu$ $e$TL and $\mu$TL channels are required to have $m_{\ell\ell} > 15\GeV$ in order to
reject backgrounds from vector-meson decays. The requirement also helps to suppress backgrounds in these channels from $b$-quark production.

\item Events in the $ee$ and $\mu\mu$ channels must satisfy $\MET>60\GeV$ and $|m_{\ell\ell} - m_Z| > 10\GeV$,
to suppress backgrounds from $\Zg$+jets and multijets.

\item
Events in the $e\mu$ channel are required to satisfy $\HT>130\GeV$. No \MET\ or $m_{\ell\ell}$\ cuts are applied.

\item The lepton+track event candidates must have  $\MET>45\GeV$, \HT\ (including the
track lepton) $>150\GeV$, and $|m_{\ell\ell} - m_Z| > 10\GeV$.

\end{itemize}

A parallel selection with the additional requirement of at least one
$b$-tagged jet is made. Because of the enhanced background rejection
afforded by the $b$-tag requirement, the selection is further
optimized, resulting in an $\MET$\ requirement for $ee$ and $\mu\mu$
events that is relaxed to $\MET>40\GeV$, while the $\HT$\
requirement for $e\mu$ events remains the same as for the
non-$b$-tag selection, i.e. $\HT> 130\GeV$.  We refer to the
analysis that requires at least one $b$-tagged jet as the `$b$-tag
analysis', and the events selected therein as the `$b$-tagged
sample'.  The subset of the $b$-tagged sample with $40\GeV<
\MET<60\GeV$ is referred to as the `exclusive $b$-tagged sample' and
has no overlap with the non-$b$-tag sample.

The acceptance times the branching fraction of $\ttbar$ to
dileptons, for the selection described above, is 0.96\% for the $ee
+ \mu\mu +e\mu$ channels without $b$-tagging, 0.11\% for the
exclusive $b$-tagged sample, and 0.19\% for $e$TL + $\mu$TL
channels.

\section{Background evaluation}
\label{s:backgrounds}

The $\ttbar$ event selection rejects $\Zg$+jets
events with $ee$ and $\mu\mu$ invariant mass below $15 \GeV$, or within $10 \GeV$ of the $Z$-boson mass. However, $\Zg$+jets events with $ee$ or $\mu\mu$  invariant mass outside of these regions can enter the signal sample when there is large $\met$, typically from mismeasurement.
These events are difficult to properly model in simulations
due to uncertainties on the non-Gaussian tails of the $\met$\ distribution,
on the cross section for $Z$~boson production with multiple jets, and on the
lepton energy resolution.

To evaluate the $\Zg$+jets background in dielectron and dimuon events
($Z \to \tau\tau$ is considered below), the MC prediction for the number of events in the signal region is normalized to the data using the number of
$\Zg$+jets events measured in a control region~\cite{ATL-CONF-2011-034}.
The control region is formed by
events with the same jet requirements as the signal region, but with
 $m_{\ell \ell}$ within 10 GeV of the $Z$-boson mass, and a $\MET$ cut of
$\met>45~\GeV$ for the lepton+track candidates and $\met>30~\GeV$
for the others. Contamination in the control region from other
physics processes (signal and other background processes considered
for the analysis) is subtracted according to MC predictions.
The ratio of data events to MC expectation in the control region
provides a scale factor that is used to correct the MC prediction
for $\Zg$+jets events in the signal region.

Other backgrounds mainly come from $W$+jets, \ttbar\, lepton+jets,
and single top-quark production with fake leptons. The term `fake
lepton' is used to refer to both misidentified and non-prompt lepton
candidates, the latter category arising from hadron decays in
flight. The yield of events with fake identified leptons is
evaluated from the data using a matrix method~\cite{top2010}. In
addition to the standard lepton selection requirements, a selection
with a looser isolation requirement is defined.  Dilepton events are
selected using the loose isolation requirement and events are
categorized according to whether each lepton passes the standard
selection or the loose selection but not the standard selection.
There are four such categories for the two leptons: loose-loose,
loose-standard, standard-loose, and standard-standard.  Each of the
four categories is related to the number of events with two `real'
(prompt) leptons, two fake leptons, or one of each, through a set of
linear equations with coefficients given by the products of
probabilities for real or fake lepton satisfying the loose selection
to also satisfy the standard selection.  These linear expressions
form a matrix that is inverted in order to extract the real and fake
lepton content of the observed dilepton event sample.
The probability for real leptons is measured as a function of jet
multiplicity using data samples of
$\Zee$ and $\Zmm$ events.
The corresponding probability for
fake leptons is measured in a data sample dominated by dijet production,
with events containing one lepton candidate passing the looser isolation cuts and having $\met <20 \GeV$.
Contributions from real leptons due to $W$+jets final states are
subtracted using simulated events.

For lepton+track events the largest background is from events with
fake leptons, dominated by fake track leptons. The
probability of a jet being reconstructed as a track lepton is
determined from a $\gamma$+jets data sample selected with photon
triggers. The fake probability is applied to a second sample
enriched in $W$+jets events with exactly one identified lepton and
no track leptons, but using the same kinematic cuts as for the
signal sample. In this second sample the fake probabilities are
summed for each jet in each event and the fake track-lepton
contribution is calculated as a function of the number of jets.

The contributions from other electroweak background processes with
two real leptons,  such as single top quarks, $Z \to \tau\tau$,
$WW$, $ZZ$ and $WZ$ production are determined from Monte Carlo
simulations.   The expected numbers of background events are given
in Table~\ref{t:signal}.  The absence of $\Zg$+jets background in
the $e\mu$ channel, coupled with the larger branching fraction for
the $e\mu$ signal, allows relatively loose selection criteria to be
used in this channel. 

The background contributions for the $b$-tag analysis are determined
using the same techniques described above, with the additional
requirement of a $b$-tagged jet.

The modeled acceptances, efficiencies and data-driven background
evaluation methods are validated by comparing predictions from Monte
Carlo simulations with data in control regions with kinematics
similar to the signal region but dominated by backgrounds. In
particular, the \met, $m_{\ell\ell}$ and jet multiplicity
distributions are studied in a sample of $Z$-boson candidates,
defined by requiring $|m_{\ell\ell} - m_Z| < 10 \GeV$ and $\met
<$~60 GeV. The predictions from MC simulations in the control
regions are in reasonable agreement with data, although small
discrepancies exist in regions that do not affect the $\ttbar$
cross-section measurement.

\newcommand{\DYZeeNJetsTwoJet}{$4.0^{+2.5}_{-1.2}$}
\newcommand{\DYZmmNJetsTwoJet}{$14.4^{+5.4}_{-4.2}$}
\newcommand{\ZtteeNJetsTwoJet}{4.9 $\pm$ 2.6}
\newcommand{\ZttmmNJetsTwoJet}{11.0 $\pm$ 5.0}
\newcommand{\ZttemNJetsTwoJet}{43 $\pm$ 16}
\newcommand{\FakeWeeNJetsTwoJet}{4.0 $\pm$ 5.0}
\newcommand{\FakeWmmNJetsTwoJet}{6.3 $\pm$ 4.1}
\newcommand{\FakeWemNJetsTwoJet}{44 $\pm$ 24}
\newcommand{\singletopeeNJetsTwoJet}{6.4$^{+1.2}_{-1.1}$}
\newcommand{\singletopmmNJetsTwoJet}{16.0$^{+1.9}_{-2.2}$}
\newcommand{\singletopemNJetsTwoJet}{41.1 $\pm$ 5.5}
\newcommand{\dibosoneeNJetsTwoJet}{5.9 $\pm$ 1.1}
\newcommand{\dibosonmmNJetsTwoJet}{8.7$^{+1.2}_{-1.5}$}
\newcommand{\dibosonemNJetsTwoJet}{32.9 $\pm$ 4.9}

\newcommand{\TotalNonttbaree}{25.2 $\pm$ 6.4}
\newcommand{\TotalNonttbarmm}{56.5 $\pm$ 9.4}
\newcommand{\TotalNonttbarem}{161 $\pm$ 34}

\newcommand{\ttbareeNJetsTwoJet}{124 $\pm$ 17}
\newcommand{\ttbarmmNJetsTwoJet}{$241^{+15}_{-18}$}
\newcommand{\ttbaremNJetsTwoJet}{746 $\pm$ 42}

\newcommand{\TotalExpectedee}{149 $\pm$ 18}
\newcommand{\TotalExpectedmm}{$298^{+17}_{-20}$}
\newcommand{\TotalExpectedem}{907 $\pm$ 54}

\newcommand{\DataeeNJetsTwoJet}{165}
\newcommand{\DatammNJetsTwoJet}{301}
\newcommand{\DataemNJetsTwoJet}{963}

\newcommand{\DYZeTLNJetsTwoJet}{$24.3^{+10.7}_{-9.4}$}
\newcommand{\DYZmTLNJetsTwoJet}{$22.0^{+5.3}_{-5.8}$}
\newcommand{\ZtteTLNJetsTwoJet}{$17.0^{+8.4}_{-7.6}$}
\newcommand{\ZttmTLNJetsTwoJet}{25 $\pm$ 11}
\newcommand{\FakeWeTLNJetsTwoJet}{74 $\pm$ 15}
\newcommand{\FakeWmTLNJetsTwoJet}{85 $\pm$ 17}
\newcommand{\singletopeTLNJetsTwoJet}{$5.7^{+1.0}_{-0.9}$}
\newcommand{\singletopmTLNJetsTwoJet}{$6.3^{+0.8}_{-1.1}$}
\newcommand{\dibosoneTLNJetsTwoJet}{$5.9^{+0.9}_{-0.8}$}
\newcommand{\dibosonmTLNJetsTwoJet}{$4.8^{+0.6}_{-0.7}$}

\newcommand{\TotalNonttbareTL}{$126^{+20}_{-19}$}
\newcommand{\TotalNonttbarmTL}{142 $\pm$ 21}

\newcommand{\ttbareTLNJetsTwoJet}{$112~^{+16}_{-18}$}
\newcommand{\ttbarmTLNJetsTwoJet}{$110~^{+17}_{-16}$}

\newcommand{\TotalExpectedeTL}{239 $\pm$ 26}
\newcommand{\TotalExpectedmTL}{253 $\pm$ 27}

\newcommand{\DataeTLNJetsTwoJet}{236}
\newcommand{\DatamTLNJetsTwoJet}{255}

\newcommand{\DYZeeNJetsTwoJetb}{$9.8^{+1.7}_{-1.3}$}
\newcommand{\DYZmmNJetsTwoJetb}{$20.3^{+1.8}_{-2.8}$}
\newcommand{\ZtteeNJetsTwoJetb}{$1.8^{+1.1}_{-1.2}$}
\newcommand{\ZttmmNJetsTwoJetb}{7.6$^{+3.3}_{-3.6}$}
\newcommand{\ZttemNJetsTwoJetb}{9.5$^{+4.2}_{-3.9}$}
\newcommand{\FakeWeeNJetsTwoJetb}{7.5 $\pm$ 6.5}
\newcommand{\FakeWmmNJetsTwoJetb}{4.9 $\pm$ 3.1}
\newcommand{\FakeWemNJetsTwoJetb}{20 $\pm$ 13}
\newcommand{\singletopeeNJetsTwoJetb}{7.3$^{+1.3}_{-1.1}$}
\newcommand{\singletopmmNJetsTwoJetb}{16.2$^{+2.2}_{-2.3}$}
\newcommand{\singletopemNJetsTwoJetb}{33.5$^{+4.8}_{-4.7}$}
\newcommand{\dibosoneeNJetsTwoJetb}{2.2 $\pm$ 0.7}
\newcommand{\dibosonmmNJetsTwoJetb}{2.6$^{+0.9}_{-0.6}$}
\newcommand{\dibosonemNJetsTwoJetb}{8.8$^{+1.7}_{-1.6}$}

\newcommand{\TotalNonttbareeb}{28.6 $\pm$ 6.9}
\newcommand{\TotalNonttbarmmb}{51.6$^{+5.6}_{-5.9}$}
\newcommand{\TotalNonttbaremb}{71.6 $\pm$ 14.1}

\newcommand{\ttbareeNJetsTwoJetb}{159$^{+17}_{-21}$}
\newcommand{\ttbarmmNJetsTwoJetb}{$304^{+26}_{-35}$}
\newcommand{\ttbaremNJetsTwoJetb}{675$^{+57}_{-75}$}

\newcommand{\TotalExpectedeeb}{188$^{+18}_{-22}$}
\newcommand{\TotalExpectedmmb}{$356^{+27}_{-35}$}
\newcommand{\TotalExpectedemb}{746$^{+59}_{-76}$}

\newcommand{\DataeeNJetsTwoJetb}{201}
\newcommand{\DatammNJetsTwoJetb}{365}
\newcommand{\DataemNJetsTwoJetb}{834}

\begin{table*}[htb]
\begin{footnotesize}
\begin{adjustwidth}{-6em}{-6em}
\centering
\begin{tabular}{|l|c|c|c|c|c||c|c|c|} \hline
                          & $ee$                    & $\mu\mu$                & $e\mu$                  & $e$TL                & $\mu$TL      & $b$-tag $ee$         & $b$-tag $\mu\mu$      & $b$-tag $e\mu$          \\ [0.2ex] \hline
                          &                         &                         &                         &                      &              &                      &                       &                         \\ [-1.9ex]
                          $\Zg$+~jets              & \DYZeeNJetsTwoJet       & \DYZmmNJetsTwoJet       & -                       & \DYZeTLNJetsTwoJet  & \DYZmTLNJetsTwoJet         & \DYZeeNJetsTwoJetb     & \DYZmmNJetsTwoJetb    & $-$                             \\ [0.3ex]
$\Zg \to \tau\tau$+jets  & \ZtteeNJetsTwoJet       & \ZttmmNJetsTwoJet       & \ZttemNJetsTwoJet       & \ZtteTLNJetsTwoJet  & \ZttmTLNJetsTwoJet                & \ZtteeNJetsTwoJetb            & \ZttmmNJetsTwoJetb     &  \ZttemNJetsTwoJetb           \\ [0.3ex]
 Fake leptons & \FakeWeeNJetsTwoJet     & \FakeWmmNJetsTwoJet     & \FakeWemNJetsTwoJet     & \FakeWeTLNJetsTwoJet          & \FakeWmTLNJetsTwoJet                   & \FakeWeeNJetsTwoJetb     & \FakeWmmNJetsTwoJetb     & \FakeWemNJetsTwoJetb           \\ [0.3ex]
Single top quark              & \singletopeeNJetsTwoJet & \singletopmmNJetsTwoJet & \singletopemNJetsTwoJet & \singletopeTLNJetsTwoJet         & \singletopmTLNJetsTwoJet         & \singletopeeNJetsTwoJetb & \singletopmmNJetsTwoJetb & \singletopemNJetsTwoJetb       \\ [0.3ex]
Diboson                  & \dibosoneeNJetsTwoJet   & \dibosonmmNJetsTwoJet   & \dibosonemNJetsTwoJet   & \dibosoneTLNJetsTwoJet          & \dibosonmTLNJetsTwoJet         & \dibosoneeNJetsTwoJetb   & \dibosonmmNJetsTwoJetb   & \dibosonemNJetsTwoJetb        \\ [0.3ex] \hline
                          &                         &                         &                         &                      &              &                      &                       &                         \\ [-1.9ex]
 Total background         & \TotalNonttbaree        & \TotalNonttbarmm        & \TotalNonttbarem        & \TotalNonttbareTL     & \TotalNonttbarmTL       & \TotalNonttbareeb        & \TotalNonttbarmmb        & \TotalNonttbaremb           \\ [0.3ex]
Predicted \ttbar\       & \ttbareeNJetsTwoJet     & \ttbarmmNJetsTwoJet     & \ttbaremNJetsTwoJet     & \ttbareTLNJetsTwoJet       & \ttbarmTLNJetsTwoJet         & \ttbareeNJetsTwoJetb     & \ttbarmmNJetsTwoJetb     & \ttbaremNJetsTwoJetb \\ [0.3ex] \hline
                          &                         &                         &                         &                      &              &                      &                       &                         \\ [-1.9ex]
Total                 & \TotalExpectedee        & \TotalExpectedmm        & \TotalExpectedem        & \TotalExpectedeTL & \TotalExpectedmTL                 & \TotalExpectedeeb        & \TotalExpectedmmb       & \TotalExpectedemb \\ [0.3ex] \hline \hline
                          &                         &                         &                         &                      &              &                      &                       &                         \\ [-1.9ex]
Observed               & \DataeeNJetsTwoJet      & \DatammNJetsTwoJet      & \DataemNJetsTwoJet      & \DataeTLNJetsTwoJet                   & \DatamTLNJetsTwoJet                                    & \DataeeNJetsTwoJetb      & \DatammNJetsTwoJetb      & \DataemNJetsTwoJetb                      \\ \hline
\end{tabular}
\end{adjustwidth}
\end{footnotesize}
\caption {Breakdown of the expected {\ttbar} signal and background
events in the signal region compared to the observed event yields,
for each of the dilepton channels. All systematic uncertainties are
included, and correlations between different background sources are
taken into account, when calculating the total background
uncertainty. The largest contribution to the line labeled 'Fake
leptons' comes from $W+$jets events.} \label{t:signal}
\end{table*}

\section{Systematic uncertainties}
\label{s:objsyst}
A summary of the systematic uncertainties on the measured $\ttbar$ production cross section is given in Table~\ref{tab:syssum}.

Lepton trigger efficiencies, and reconstruction and selection efficiencies for identified leptons and track leptons, are assessed using
$Z\rightarrow ee$ and $Z\rightarrow\mu\mu$ events in the same data
sample as used for the $\ttbar$ analyses.
Scale factors are evaluated by comparing these efficiencies with those determined with simulated $Z$-boson events. The scale factors are applied to MC samples when calculating
acceptances to account for any differences between predicted and observed efficiencies.  Systematic uncertainties on these scale factors are evaluated by varying the selection of events used in the efficiency measurements and by checking the stability of the measurements over the course of the data-taking period.

The modeling of lepton momentum scale and resolution is studied using
reconstructed dilepton invariant mass distributions of $\Zg$\ candidate events, and
the simulation is adjusted to match the data.  Uncertainties in the scale and resolution are used to
evaluate the systematic uncertainty due to these corrections.
The acceptance uncertainty from the lepton modeling is dominated
mostly by the electron selection efficiency uncertainty.

The jet energy scale (JES) and its uncertainty are derived by
combining information from test-beam data, LHC collision data and
simulation~\cite{jetcor}. For jets within the acceptance, the JES
uncertainty varies in the range 4--8\% as a function of jet $\pT$
and $\eta$. This uncertainty is higher than in the previous
result~\cite{ATL-CONF-2011-034} because of the additional
uncertainty due to multiple $pp$ interactions at high instantaneous
luminosity. The jet energy resolution and jet
reconstruction/identification efficiency measured in data and in
simulation are in good agreement. The statistical uncertainties on
the comparisons,  10\%\ and 1--2\%\ for the energy resolution and
the efficiency, respectively, are taken as systematic uncertainties
associated with these effects. The effect on the acceptance is
dominated by the JES uncertainty.

The systematic uncertainty on the efficiency of the $b$-tagging
algorithm has been estimated to be $6$\%\ for $b$-quark jets, based
on $b$-tagging calibration studies using inclusive lepton and
multijet final states. The uncertainties on the tagging rates of
jets from light and charm quarks are larger, but are not a
significant source of uncertainty due to the intrinsically high
signal-to-background ratios in the dilepton final states. The
acceptance uncertainty due to $b$-tagging is about 3\% for all three
channels.

The uncertainty in the kinematic distributions of the \ttbar\ signal events
gives rise to systematic uncertainties in
the signal acceptance, with contributions from the choice of generator, the modeling of initial- and final-state
radiation (ISR/FSR) and the PDFs.
The generator and parton-showering uncertainty (collectively labeled
`Generator' in Table~\ref{tab:syssum}) are evaluated by comparing the
{\sc MC@NLO} predictions  with those of
{\sc Powheg}~\cite{powheg,Frixione:2007vw,Alioli:2010xd} interfaced to
either {\sc Herwig} or {\sc Pythia}.
The uncertainty due to ISR/FSR is evaluated using the {\sc AcerMC} generator~\cite{Acer}
interfaced to the {\sc Pythia}\ shower model, and by varying the parameters controlling ISR
and FSR in a range consistent with those used in the Perugia Hard/Soft tune variations~\cite{Skands}.
Finally, the PDF uncertainty is evaluated using a range of
current PDF sets~\cite{cteq6l}.
The dominant uncertainties in this category of systematics are the
modeling of ISR/FSR and the generator choice.

The overall normalization uncertainties on the backgrounds from
single top quark and diboson production are taken to be
8.6\%~\cite{PhysRevD.83.091503} and 5\%~\cite{Campbell:2010ff},
respectively.
The systematic uncertainties from the background evaluations derived
from the data include the statistical uncertainties in these methods
as well as the systematic uncertainties arising from lepton and jet
identification and reconstruction, and the MC estimates that are
used. An uncertainty on the data-driven $\Zg$+jets evaluation, based
on the expected $\met$ resolution in these events, is included by
varying the $\met$ cut in the control region by $\pm 5\GeV$ for $ee$
and $\mu\mu$ events, and $\pm 10\GeV$ in lepton+track events where
the $\met$ resolution is poorer. The systematic uncertainty on the
fake identified lepton background prediction is 50\%, as measured
using control regions with different flavor composition and photon
conversion rates, as determined by Monte Carlo studies. A 20\%
systematic uncertainty is set on the prediction of the fake
track-lepton background, derived from a comparison of predicted and
observed fake track leptons in control regions defined as
opposite-sign events with zero or one jet without an $\HT$ cut, and
same-sign events with more than one jet.
The uncertainty on the measured integrated luminosity of the dataset
is 3.7\%~\cite{lumi}.

Table~\ref{tab:syssum} lists the contributions to the cross-section
measurement of each of the systematic uncertainties considered, in
percent, with the non-$b$-tag and the exclusive $b$-tag analyses
combined in the $ee$ and $\mu\mu$ columns.  Only non-$b$-tag events
are used in the $e\mu$ channel.  The relatively large `Generator'
uncertainty for $ee$ events is partly a result of the limited size
of the MC data set.  In the $e$TL and $\mu$TL columns the `MC
statistics' uncertainty is larger than in the $ee$, $\mu\mu$, and
$e\mu$ case because the number of events available is reduced by the
removal of events with two identified leptons.  The combined
uncertainty comes from the profile likelihood technique used to
determine the cross section~\cite{ATL-CONF-2011-034}, and takes into
account all correlations.  The statistical uncertainty is determined
by fixing all systematic uncertainties at their best-fit values in
the likelihood function.

%
\begin{table*}[htb]
\begin{footnotesize}
\begin{adjustwidth}{-6em}{-6em}
\centering
\begin{tabular}{|l|c|c|c|c|c|c|}
\hline
           Uncertainties       $\Delta\sigma/\sigma$[\%]        & $ee$              & $\mu\mu$          & $e\mu$            &  $e$TL  & $\mu$TL   &Combined         \\ \hline


Data statistics           &    $\pm 8.1$  &    $\pm 6.1$  &    $\pm 3.9$  &   $\pm 14.1$  &   $\pm 14.2$  &    $\pm 2.9$\\
\hline
Luminosity                &    +4.4/-3.8 &    +4.4/-3.9  &    $\pm 4.2$  &    +5.1/-4.2  &    +5.4/-4.4  &    $\pm 4.3$ \\
\hline
MC statistics             &    $\pm 1.6$  &    $\pm 1.2$  &    $\pm 0.8$  &    $\pm 5.5$  &    $\pm 4.6$  &    +0.7/-0.6 \\
Lepton uncertainties      &    +6.2/-5.4   &    +2.9/-1.3  &    $\pm 3.1$  &    $\pm 4.1$  &    +1.8/-1.6  &    +2.6/-2.2 \\
Track leptons             &     ---          &    ---          & ---          &    $\pm 4.4$  &    $\pm 1.9$  &    +0.3/-0.2 \\
Jet/$\met$ uncertainties  &    +5.7/-5.7  &    +6.4/-3.5  & +4.7/-3.2  &    +14.8/-6.4  &   $\pm 13.1$  &    +4.4/-3.4 \\
$b$-tagging uncertainties &    +1.2/-1.0  &    $\pm 0.7$   &     ---          &     ---          &     ---          &    +0.4/-0.0 \\
$\Zg$+~jets evaluation    &    $\pm 0.4$  &    +0.5/-0.0  &      --- &    $\pm 6.2$  &    +2.4/-2.7  &    +0.3/-0.2 \\
Fake lepton evaluation    &    $\pm 3.3$  &    1.5/-1.3  &    $\pm 3.0$  &   $\pm 13.7$  &   $\pm 15.1$  &    $\pm 1.7$  \\
Generator                 &   +12/-11     &    +4.5/-4.3   &    +4.8/-4.5   &   +14/-11    &   +14/-13     &    +5.1/-4.9  \\
\hline
All syst.(except lumi.)    &   +16.4/-14.4  &    +8.8/-6.4  &    +8.2/-6.8  &   +27.9/-20.7  &   +26.5/-23.7   &    +8.0/-6.5 \\
\hline \hline
Stat. + syst.              &   +18.9/-16.9  &    +11.6/-9.5  &    +10.1/-8.8  &   +31.8/-25.2  &   +30.7/-27.8  &    +9.6/-8.2  \\
\hline

\end{tabular}
\caption{Overview of the \ttbar{} cross-section uncertainties.}
\label{tab:syssum}
\end{adjustwidth}
\end{footnotesize}
\end{table*}

\section{Cross-section measurement}
\label{s:nobtagging}

The expected and measured numbers of events in the signal region,
after applying all selection cuts for each of the individual
dilepton channels, are shown in Table~\ref{t:signal}. A total of
1920 candidate events are observed for the analysis without
$b$-tagging, and a total of 1400 candidate events are found for the
$b$-tag analysis.  There are 1221 events in common between the two
selections, and 179 exclusive $b$-tagged events.

\begin{figure*}
  \centering
   \subfigure[]{
    \includegraphics[width=0.45\textwidth]{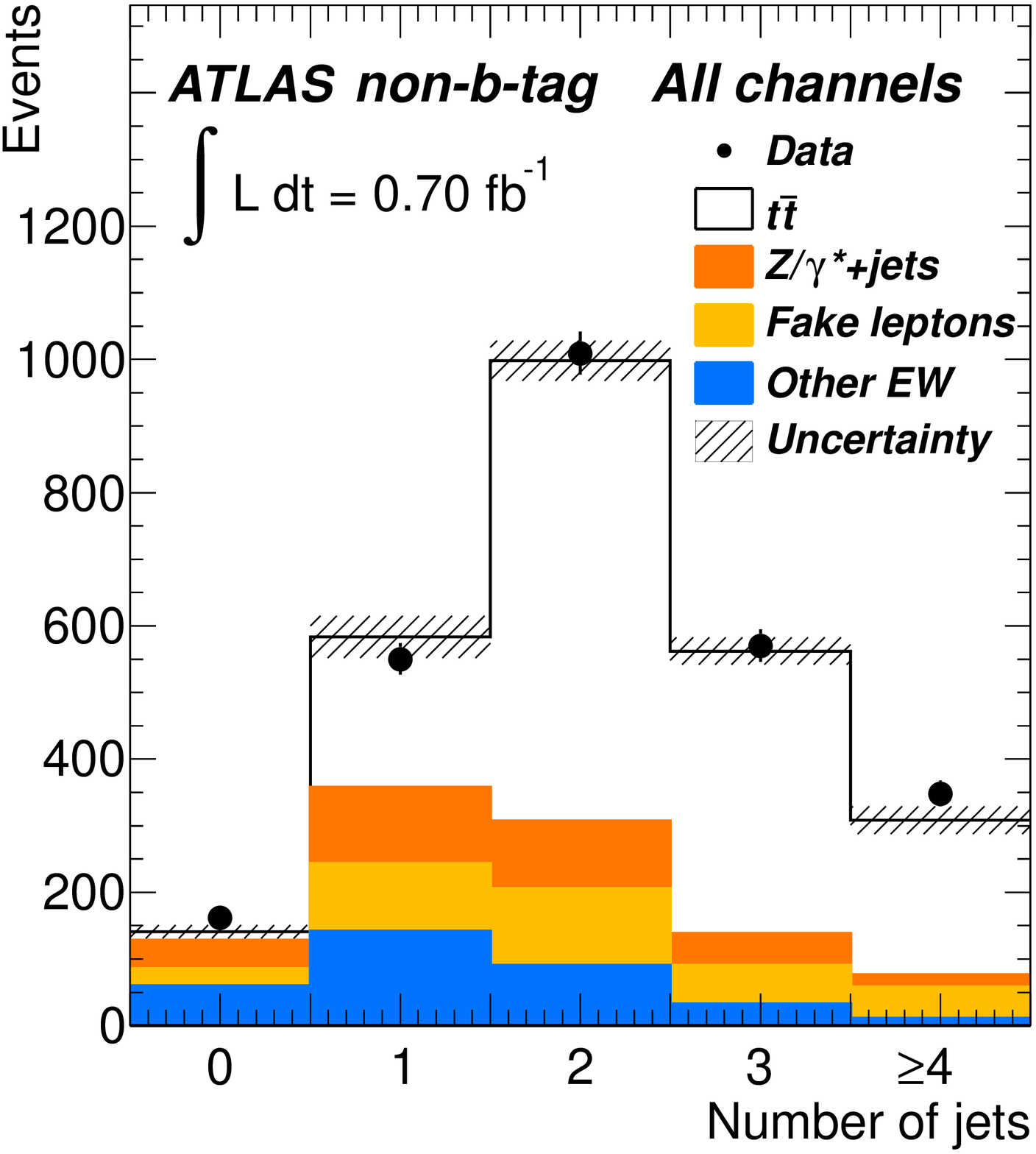}
    }
   \subfigure[]{
    \includegraphics[width=0.45\textwidth]{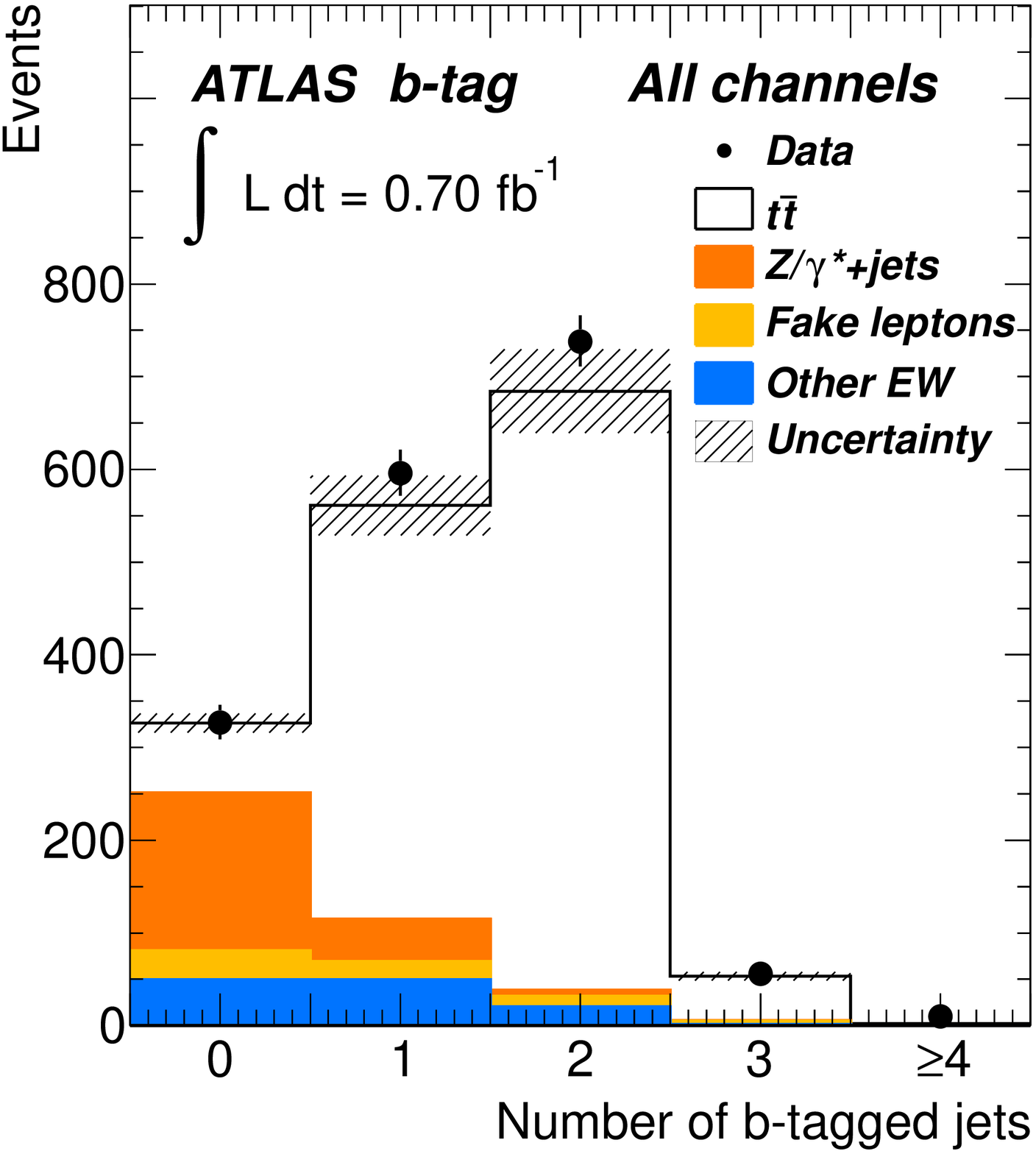}
    }
  \caption{ (a) Jet multiplicity distribution for
$ee$+$\mu\mu$+$e\mu$+$e$TL+$\mu$TL events without a $b$-tagging requirement. (b)
 Multiplicity distribution of $b$-tagged jets in the $ee$+$\mu\mu$+$e\mu$ channels. Contributions from diboson and
single top-quark events are summarized as `Other EW'. The
events in (b) are not a simple subset of those in (a) because the
event selections for the $b$-tag and non-$b$-tag analyses differ.
Uncertainties shown are statistical and systematic combined. The
distributions are shown as stacked histograms.}
  \label{f:ll_njets}
\end{figure*}

\begin{figure*}
  \centering
     \subfigure[]{
    \includegraphics[width=0.41\textwidth]{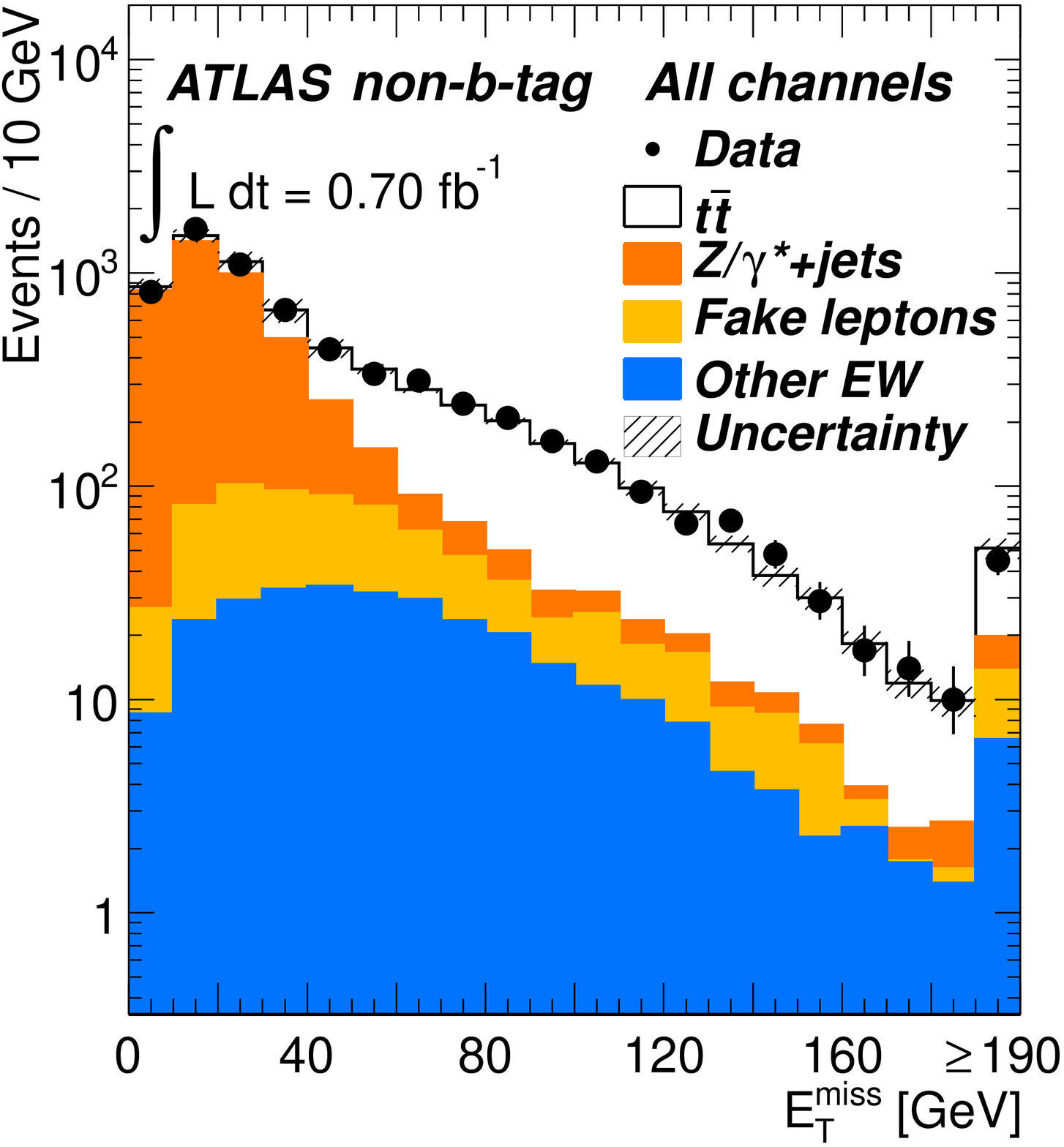}
    }
   \subfigure[]{
    \includegraphics[width=0.41\textwidth]{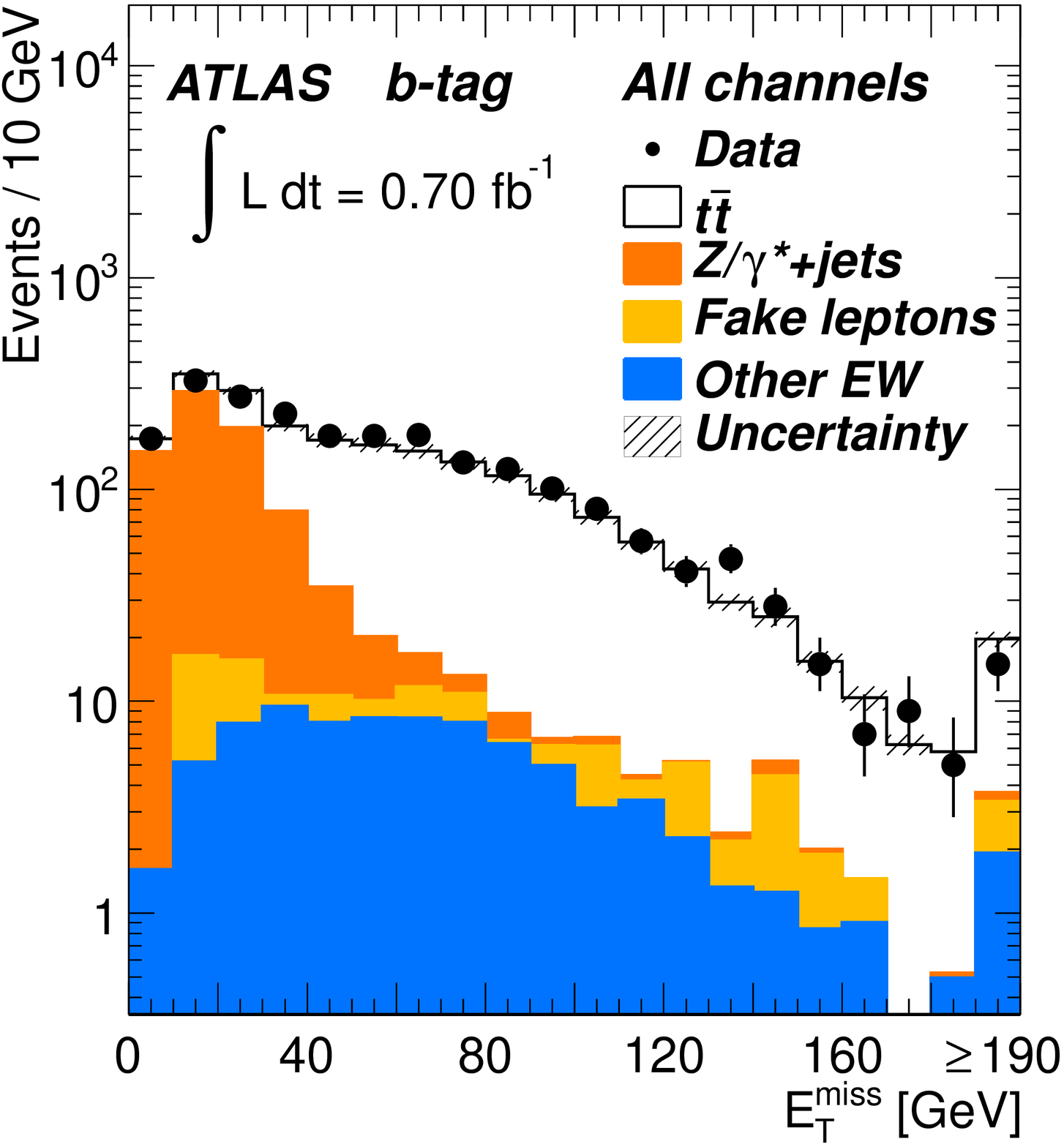}
    }
  \caption{
The $\met$ distribution in the signal region for (a) the five
non-$b$-tag channels combined and, (b) the three $b$-tagged channels
combined.  Contributions from diboson and single top-quark events
are summarized as `Other EW'. Uncertainties shown are statistical
and systematic combined. The last bin in each figure is an
overflow bin, including all events above 190 GeV. The distributions
are shown as stacked histograms.}
  \label{f:ll_met_ht}
\end{figure*}

In Fig.~\ref{f:ll_njets} the number of selected jets and the
expectation for $\lumitot$ are shown for the non-$b$-tag analysis
with the five channels combined, and for the $b$-tag analysis with
the three channels combined. In the non-$b$-tag case, all
requirements except the jet multiplicity selection are applied, and
in the $b$-tag case all requirements except the $b$-tag requirement
are applied. The $\met$ distributions for the combination of the
five non-$b$-tag and three $b$-tagged channels are
shown in Fig.~\ref{f:ll_met_ht}. All requirements except $\met$ are
applied.
The dominant backgrounds are $\Zg +$jets and $W$+jets
production with a fake lepton and, for the $b$-tag analysis,
single-top events.

The cross-section results are obtained with a profile likelihood
technique, as described in Ref.~\cite{ATL-CONF-2011-034}. The
branching fraction for $t\rightarrow Wb$ is taken to be 100\% and
the acceptance is calculated for a top mass of 172.5 GeV.

\begin{figure}[!h]
\centering
\includegraphics[width=0.5\textwidth]{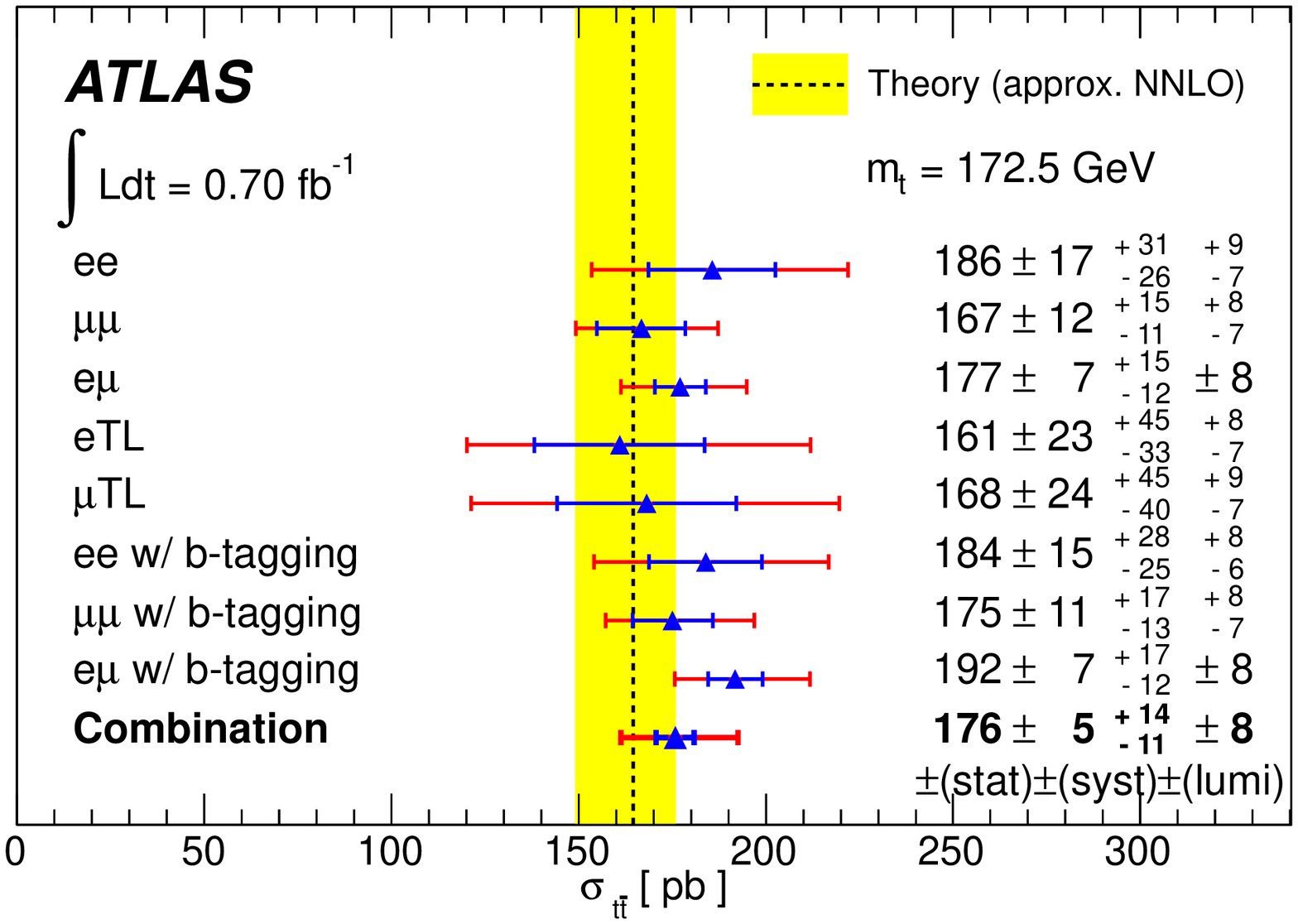}
    \caption{Summary of the individual cross section measurements and the combination of non-$b$-tag and
    exclusive $b$-tagged results. The vertical dashed line and yellow band are the approximate NNLO theory calculation and its uncertainty.}
    \label{fig:XsecSummary}
\end{figure}

The top-quark pair production cross section measured by combining
the seven channels, the non-$b$-tagged $ee$, $\mu\mu$, $e\mu$, $e$TL
and $\mu$TL and the exclusive $b$-tagged $ee$ and $\mu\mu$, is
\[
\sigmattbar=\mathrm{}\xsecctot \xseccstat
\mathrm{(stat.)}\xseccsyst \mathrm{(syst.)} \xsecclumi
\mathrm{(lumi.)}~\rm pb.
\]
Figure~\ref{fig:XsecSummary} summarizes the cross sections
for the individual channels,  and the combination of the non-$b$-tag and
the exclusive $b$-tagged data sets.

The measured cross section is in good agreement with a similar
measurement made with 2010 data by the CMS
collaboration~\cite{Chatrchyan:2011nb}, 
with an ATLAS measurement in the dilepton channel with earlier
data~\cite{ATL-CONF-2011-034}, and with the SM
prediction of
$165\pmasym{11}{16}$~pb. Compared to the earlier ATLAS measurement
in the dilepton channel, the statistical uncertainty of the
measurement has been reduced by a factor of four with the addition
of more data, and a small reduction in the systematic uncertainty,
which now dominates, has been achieved.


\section{Acknowledgements}

We thank CERN for the very successful operation of the LHC, as well as the
support staff from our institutions without whom ATLAS could not be
operated efficiently.

We acknowledge the support of ANPCyT, Argentina; YerPhI, Armenia; ARC,
Australia; BMWF, Austria; ANAS, Azerbaijan; SSTC, Belarus; CNPq and FAPESP,
Brazil; NSERC, NRC and CFI, Canada; CERN; CONICYT, Chile; CAS, MOST and NSFC,
China; COLCIENCIAS, Colombia; MSMT CR, MPO CR and VSC CR, Czech Republic;
DNRF, DNSRC and Lundbeck Foundation, Denmark; EPLANET and ERC, European Union;
IN2P3-CNRS, CEA-DSM/IRFU, France; GNAS, Georgia; BMBF, DFG, HGF, MPG and AvH
Foundation, Germany; GSRT, Greece; ISF, MINERVA, GIF, DIP and Benoziyo Center,
Israel; INFN, Italy; MEXT and JSPS, Japan; CNRST, Morocco; FOM and NWO,
Netherlands; RCN, Norway; MNiSW, Poland; GRICES and FCT, Portugal; MERYS
(MECTS), Romania; MES of Russia and ROSATOM, Russian Federation; JINR; MSTD,
Serbia; MSSR, Slovakia; ARRS and MVZT, Slovenia; DST/NRF, South Africa;
MICINN, Spain; SRC and Wallenberg Foundation, Sweden; SER, SNSF and Cantons of
Bern and Geneva, Switzerland; NSC, Taiwan; TAEK, Turkey; STFC, the Royal
Society and Leverhulme Trust, United Kingdom; DOE and NSF, United States of
America.

The crucial computing support from all WLCG partners is acknowledged
gratefully, in particular from CERN and the ATLAS Tier-1 facilities at
TRIUMF (Canada), NDGF (Denmark, Norway, Sweden), CC-IN2P3 (France),
KIT/GridKA (Germany), INFN-CNAF (Italy), NL-T1 (Netherlands), PIC (Spain),
ASGC (Taiwan), RAL (UK) and BNL (USA) and in the Tier-2 facilities
worldwide.

\bibliographystyle{atlasnote}
\bibliography{DileptonSigmaJHEP}

\clearpage
\begin{flushleft}
{\Large The ATLAS Collaboration}

\bigskip

G.~Aad$^{\rm 48}$,
B.~Abbott$^{\rm 110}$,
J.~Abdallah$^{\rm 11}$,
A.A.~Abdelalim$^{\rm 49}$,
A.~Abdesselam$^{\rm 117}$,
O.~Abdinov$^{\rm 10}$,
B.~Abi$^{\rm 111}$,
M.~Abolins$^{\rm 87}$,
O.S.~AbouZeid$^{\rm 157}$,
H.~Abramowicz$^{\rm 152}$
% ...
\bigskip

% ...
$^{10}$ Institute of Physics, Azerbaijan Academy of Sciences, Baku, Azerbaijan\\
$^{11}$ Institut de F\'isica d'Altes Energies and Departament de F\'isica de la Universitat Aut\`onoma  de Barcelona and ICREA, Barcelona, Spain\\
% ...
$^{48}$ Fakult\"{a}t f\"{u}r Mathematik und Physik, Albert-Ludwigs-Universit\"{a}t, Freiburg i.Br., Germany\\
$^{49}$ Section de Physique, Universit\'e de Gen\`eve, Geneva, Switzerland\\
% ...
$^{87}$ Department of Physics and Astronomy, Michigan State University, East Lansing MI, United States of America\\
% ...
$^{110}$ Homer L. Dodge Department of Physics and Astronomy, University of Oklahoma, Norman OK, United States of America\\
$^{111}$ Department of Physics, Oklahoma State University, Stillwater OK, United States of America\\
% ...
$^{117}$ Department of Physics, Oxford University, Oxford, United Kingdom\\
% ...
$^{152}$ Raymond and Beverly Sackler School of Physics and Astronomy, Tel Aviv University, Tel Aviv, Israel\\
% ...
$^{157}$ Department of Physics, University of Toronto, Toronto ON, Canada\\
% ...
\end{flushleft}


\clearpage

\end{document}